\documentclass[twocolumn,epjc3]{svjour3}

\usepackage[english]{babel}  
\usepackage[utf8]{inputenc}
\usepackage[T1]{fontenc}
\usepackage{amsmath, amssymb}
\usepackage{hyperref}
\usepackage{xfrac}
\usepackage[dvipsnames]{xcolor}
\usepackage[justification=centering]{caption, subcaption}
\usepackage{float}

\hypersetup{
  colorlinks = true, 
  urlcolor = RedViolet, 
  linkcolor = RoyalBlue, 
  citecolor = ForestGreen 
}

\journalname{Eur. Phys. J. C}

\begin{document}

\title{Measurements of the ionization efficiency of protons in methane}

\author{L.~Balogh\thanksref{inst0} 
\and C.~Beaufort\thanksref{inst2, email} 
\and A.~Brossard\thanksref{inst1} 
\and J.-F.~Caron\thanksref{inst0} 
\and M.~Chapellier\thanksref{inst1} 
\and J.-M.~Coquillat\thanksref{inst1} 
\and E.~C.~Corcoran\thanksref{inst3} 
\and S.~Crawford\thanksref{inst1} 
\and A.~Dastgheibi-Fard\thanksref{inst2} 
\and Y.~Deng\thanksref{inst4} 
\and K.~Dering\thanksref{inst1} 
\and D. ~Durnford\thanksref{inst4} 
\and C.~Garrah\thanksref{inst4} 
\and G.~Gerbier\thanksref{inst1} 
\and I.~Giomataris\thanksref{inst5} 
\and G.~Giroux\thanksref{inst1} 
\and P.~Gorel\thanksref{inst6}
\and M.~Gros\thanksref{inst5} 
\and P.~Gros\thanksref{inst1} 
\and O.~Guillaudin\thanksref{inst2} 
\and E.~W.~Hoppe\thanksref{inst7}
\and I.~Katsioulas\thanksref{inst8}
\and F.~Kelly\thanksref{inst3} 
\and P.~Knights\thanksref{inst8}
\and S.~Langrock\thanksref{inst6} 
\and P.~Lautridou\thanksref{inst9} 
\and I.~Manthos\thanksref{inst8}
\and R.~D. Martin\thanksref{inst1}
\and J.~Matthews\thanksref{inst8}
\and J.-P.~Mols\thanksref{inst5}
\and J.-F.~Muraz\thanksref{inst2} 
\and T.~Neep\thanksref{inst8} 
\and K.~Nikolopoulos\thanksref{inst8} 
\and P.~O'Brien\thanksref{inst4}
\and M.-C.~Piro\thanksref{inst4} 
\and D.~Santos\thanksref{inst2}
\and G.~Savvidis\thanksref{inst1}
\and I.~Savvidis\thanksref{inst10}
\and F.~Vazquez de Sola Fernandez\thanksref{inst9}
\and M.~Vidal\thanksref{inst1}
\and R.~Ward\thanksref{inst8}
\and M.~Zampaolo\thanksref{inst2}
}

\institute{Department of Mechanical \& Materials Engineering, Queen's University, Kingston, Ontario K7L 3N6, Canada \label{inst0}
\and LPSC, Universit\'{e} Grenoble-Alpes, CNRS/IN2P3, Grenoble, France \label{inst2}
\and Department of Physics, Engineering Physics \& Astronomy, Queen's University, Kingston, Ontario K7L 3N6, Canada \label{inst1}
\and Chemistry \& Chemical Engineering Department, Royal Military College of Canada, Kingston, Ontario K7K 7B4, Canada \label{inst3} 
\and Department of Physics, University of Alberta, Edmonton, Alberta, T6G 2R3, Canada \label{inst4}
\and IRFU, CEA, Universit\'{e} Paris-Saclay, F-91191 Gif-sur-Yvette, France \label{inst5} 
\and SNOLAB, Lively, Ontario, P3Y 1N2, Canada \label{inst6}
\and Pacific Northwest National Laboratory, Richland, Washington 99354, USA \label{inst7}
\and School of Physics and Astronomy, University of Birmingham, Birmingham B15 2TT United Kingdom \label{inst8}
\and SUBATECH, IMT-Atlantique/CNRS-IN2P3/Nantes University, Nantes 44307, France \label{inst9}
\and Aristotle University of Thessaloniki, Thessaloniki, Greece\label{inst10}
\newline\newline \textit{NEWS-G Collaboration}}

\thankstext{email}{Corresponding author: \href{mailto:beaufort@lpsc.in2p3.fr}{beaufort@lpsc.in2p3.fr}}

\date{}

\maketitle


\begin{abstract}
The amount of energy released by a nuclear recoil ionizing the atoms of the active volume of detection appears "quenched" compared to an electron of the same kinetic energy. This different behavior in ionization between electrons and nuclei is described by the Ionization Quenching Factor (IQF) and it plays a crucial role in direct dark matter searches. For low kinetic energies (below $50~\mathrm{keV}$), IQF measurements deviate significantly from common models used for theoretical predictions and simulations. We report measurements of the IQF for proton, an appropriate target for searches of Dark Matter candidates with a mass of approximately $1~\rm{GeV}$, with kinetic energies in between $2~\mathrm{keV}$ and $13~\mathrm{keV}$ in $100~\mathrm{mbar}$ of methane. We used the Comimac facility in order to produce the motion of nuclei and electrons of controlled kinetic energy in the active volume, and a NEWS-G SPC to measure the deposited energy. The Comimac electrons are used as a reference to calibrate the detector with 7 energy points. A detailed study of systematic effects led to the final results well fitted by $\mathrm{IQF}~(E_K)= E_K^\alpha~/~(\beta + E_K^\alpha)$ with $\alpha = 0.70\pm0.08$ and $\beta = 1.32\pm0.17$. In agreement with some previous works in other gas mixtures, we measured less ionization energy than predicted from \texttt{SRIM} simulations, the difference reaching $33\%$ at $2~\mathrm{keV}$.

\end{abstract}

\begin{figure*}
\begin{minipage}{0.49\linewidth}
	\centering
	\includegraphics[width=\linewidth, height=5.9cm]{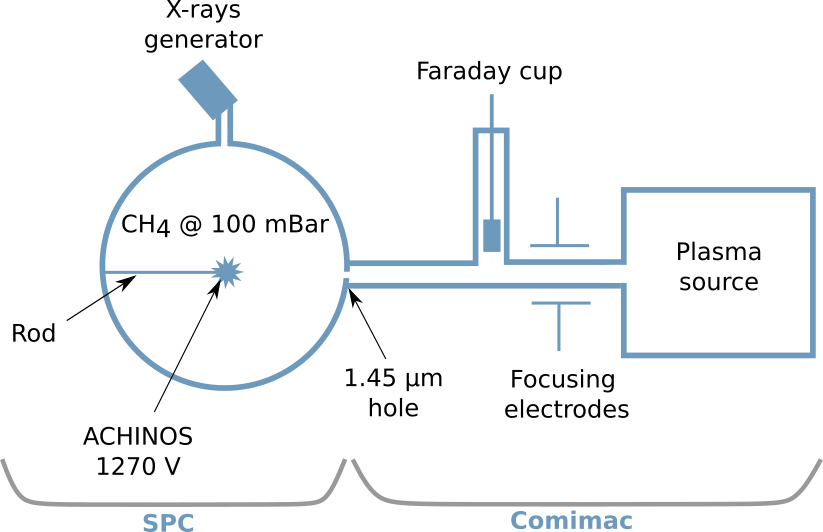}
	\caption{Experimental setup}
	\label{fig:setup}
\end{minipage}
\hfill
\begin{minipage}{0.49\linewidth}
	\centering
	\includegraphics[width=\linewidth, height=5.9cm]{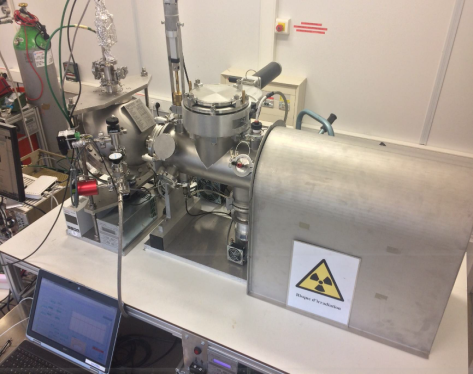}
	\caption{A picture of the experiment with Comimac on the right and the SPC on the left}
	\label{fig:picture}
\end{minipage}
\end{figure*}

\section{Introduction}

The amount of energy released as ionization by a particle passing through a medium plays an important role in multiple fields from  radiotherapy and microdosimetry \cite{Gambarini2019,Herve2021,Santa2016} to cosmology and particle detection \cite{Sciolla20009,Blum2008}.  While there is mounting evidence for Dark Matter (DM) from its gravitational effects (see \cite{Bertone} for a recent review), its nature remains unknown. Extensive searches are carried out, especially regarding low-mass Weakly Interacting Massive Particles (WIMPs) for which detection sensitivity is continuously improving \cite{APPEC}. The strategy of direct search for WIMPs relies on the detection of nuclear recoils induced by elastic scattering with WIMPs. Those rare events for low-mass WIMPs produce small ionization energies, in the keV-range or below, which are challenging to detect, hence the importance of determining precisely the energy deposited by a nuclear recoil in a medium.

A nuclear recoil or an ion moving in a medium will release its energy through three competing processes: (1) the production of electron-ion pairs called ionization, (2) scintillation, in which photons are emitted in the de-excitation of quasi-molecular states, and (3) heat produced by the motion of the nuclei and the electrons. Detectors usually only have access to one or two of these processes and the total kinetic energy of the recoil needs to be inferred, for instance by converting the observed ionization into kinetic energy. The Ionization Quenching Factor (IQF) represents the fraction of energy released as ionization by a nuclear recoil or a moving ion. Hereafter, "ion" refers equally to charged (from an ion source) or neutral (from nuclear recoils) moving atoms; despite the difference in their original charge, they are assumed to have the same IQF \cite{SRIM} due to the continuously changing state of charge during the ionization process. The IQF can be expressed as:
\begin{equation}
	\mathrm{IQF}~(E_K) ~=~ \frac{E_{\mathrm{ioniz}}}{E_K}
	\label{eq:IQFdef}
\end{equation}
where $E_K$ is the kinetic energy of the ion and $E_{\mathrm{ioniz}}$ is the deposited ionization energy. The IQF depends on physical properties such as the medium, the nucleus mass, and its kinetic energy. It must then be determined for each detection configuration in order to reconstruct the kinetic energy from measurements of the ionization energy. Measuring the IQF in the keV-range is challenging since the kinetic energy of the ion must be precisely established. Most of the experimental setups rely on neutron sources to induce recoils \cite{IQF_Ge,IQF_CDMS,IQF_CSI,Izraelevitch_2017,IQF_Ar,Marie} with coincident detection of the recoil and the scattered neutron.

The quenching effects were first theoretically described by the Lindhard theory \cite{Lindhard} which models the stopping power of particles in matter under a semiclassical approximation treating the collisions as point-like interactions. While complementary models have been developed on this basis (see \cite{SRIM} for details about several models), they do not describe low-energy ions for which the semiclassical approximation usually fails \cite{Bezrukov,Sorensen}. For energies below $50~\mathrm{keV}$, deviations from the predictions of the Lindhard theory or simulation toolkits have been measured in multiple media: in gases \cite{Marie,Olivier,IQFSantos,Tampon,TheseQuentin,TheseDonovan,Hitachi}, in noble liquids \cite{Bezrukov,Mei}, and in crystals \cite{Izraelevitch_2017,Chavarria_2016,Sarkis2020}. 

In this work, we present measurements of the IQF of the proton in $100~\mathrm{mbar}$ of methane (CH$_4$) in between $2~\mathrm{keV}$ and $13~\mathrm{keV}$ showing significant deviations from numerical and theoretical predictions. We take advantage of the Comimac facility \cite{Comimac}, a table-top particle accelerator developed for IQF measurements, to directly send ions into a detector and to precisely calibrate the detector. The gas and the working conditions have been adapted to the requirements of the NEWS-G collaboration which searches for light dark matter and coherent elastic neutrino-nucleus scattering (CE$\nu$NS) with a sub-keV sensitive detector \cite{NEWSGresults,Gerbier}. The collaboration dedicates attention to methane, a gas well-adapted to low-mass WIMP searches both because hydrogen is the lightest atom, enhancing the energy recoil due to kinematics, and because it is sensitive to spin-dependent WIMP-nucleus interactions. At the time of writing, the NEWS-G collaboration finalises the installation of a Spherical Proportional Counter (SPC) of $140~\mathrm{cm}$ diameter at SNOLAB, one of the deepest underground laboratory in the world \cite{SNOLAB}. 

The paper is organized as follows: in Section \ref{sec:setup} we describe the experimental setup. Section \ref{sec:syst} is dedicated to the study of systematic uncertainties. The calibration is presented in Section \ref{sec:calib}, while the IQF results are discussed in Section \ref{sec:results}.

\section{Experimental setup}\label{sec:setup}
    
 The experimental setup is presented in Figures \ref{fig:setup} and \ref{fig:picture}. The SPC used in the experiment consists of a grounded $30~\mathrm{cm}$ diameter vessel made of stainless steel filled with $100~\mathrm{mbar}$ of CH$_4$. A small sensor is placed at the center of the sphere supported by a grounded metallic rod and the sensor is biased to a positive high voltage ($1270~\mathrm{V}$). The sensor, called ACHINOS, presents specific features to increase the magnitude of the electric field at large radii and to operate the detector at high gain \cite{Giganon2017,Katsioulas2020,Giomataris2020}. ACHINOS consists of 11 spherical anode balls of $1~\mathrm{mm}$ diameter, the balls being uniformly distributed around a central sphere. As a first approximation, the electric field in the volume evolves as $1/r^2$ with $r$ the distance to the center, dividing the inner volume into two regions: (1) the drift region with low electric field $\sim\mathcal{O}(\mathrm{10~V/cm})$ in which the primary electrons from ionization will drift towards the sensor within typically $\sim100~\mathrm{\mu s}$ ; (2) the amplification region with $E\sim\mathcal{O}(\mathrm{10~kV/cm})$ in which the electrons are sufficiently accelerated to trigger a Townsend avalanche with gain typically $\mathcal{O}(10^4)$. The motion of secondary charges induces a current on the ACHINOS sensor. A single charge preamplifier reads the 11 anodes and the signal is digitized at $1.04~\mathrm{MHz}$. The radius of the primary interaction can be inferred from the timing properties of the measured pulse, enabling localisation and background rejection \cite{Bougamont,Savvidis}. The high sensitivity to low-energy electrons makes possible to measure a single electron response \cite{SER}. 

The SPC is coupled to the Comimac facility which is a source of electrons and ions of controlled kinetic energy in the range $[700~\mathrm{eV}~,~50~\mathrm{keV}]$. Comimac uses an Electron Cyclotron Resonance (ECR) ion source of type Comic \cite{Sortais2010} with a He/H$_2$ gas mixture. The mixture is excited by $2.45~\mathrm{GHz}$ micro-waves and turns into a plasma. The He is acting as a carrier gas to efficiently break the H$_2$ molecule into protons, as detailed in Section \ref{sec:results}. The polarity of the extraction voltage $V_{ex}$ selects either electrons or ions. Its value sets their kinetic energy by the relation $E_K = qV_{ex}$. The particle beam is then focused to a $1.45~\mathrm{\mu m}$ hole that interfaces Comimac with the SPC's surface in a way that ensures pressure independence between the SPC gas ($100~\mathrm{mbar}$) and the Comimac line ($\sim 10^{-5}~\mathrm{mbar}$). The pressure in the SPC is monitored continuously with a pressure gauge. An ion sent by Comimac mimics the motion of a nuclear recoil induced by a collision with a WIMP.

The SPC is also coupled to an X-ray generator that sends high-energy X-rays on thin layers of cadmium and aluminium. The excited atoms will emit fluorescence X-rays ($1.49~\mathrm{keV}$ for $K_x$-Al ; $3.23~\mathrm{keV}$ for $L_x$-Cd \cite{ENSDF}) that will produce photoelectrons of known energy at any radius inside the sphere. As we will discuss in Section~\ref{sec:syst} the fluorescence X-rays are used to cross-check the kinetic energy of the electrons sent by Comimac.
    
In this work, we present data for electrons in the range $[1.5~, ~13]~\mathrm{keV}$ and for protons in  $[2~, ~13]~\mathrm{keV}$. The number of events in the electron runs is always larger than $30\times 10^3$ while for protons it varies between $3\times 10^3$ and $13\times 10^3$, the disparity depending on the proportion of heavier species sent by Comimac. In this paper, we define the measured ionization energy in ADU (Analog-Digital Units) as the amplitude of the signal outputted by the charge-sensitive preamplifier (CSP) after correction of the ballistic deficit. This ballistic deficit depends on the time constant of the CSP (measured to be $51.8~\mathrm{\mu s}$) which is small compared to the signal duration of a few hundreds of microseconds. In other words, it means that the signal starts decaying before all the charges have been collected. We correct the ballistic deficit by deconvolving the CSP self-discharge following the approach described in \cite{Gavrilyuk}.

The data used for the electron calibration and the IQF measurements have been taken in a single day with the same charge of gas. The detector was stable during the measurements: the measured energy of a reference Comimac point ($6~\mathrm{keV}$ electrons) varied by less than 1\% between the beginning and the end of the experiment.


\section{Study of systematic effects}\label{sec:syst}

The measurements of the IQF represent an experimental challenge and the systematic effects must be carefully studied. In this section, we detail the main systematic effects and the dedicated analyses we have performed in order to quantify them. 

\subsection{Cosmic background}

\begin{figure}[H]
    \centering
    \includegraphics[width=\linewidth]{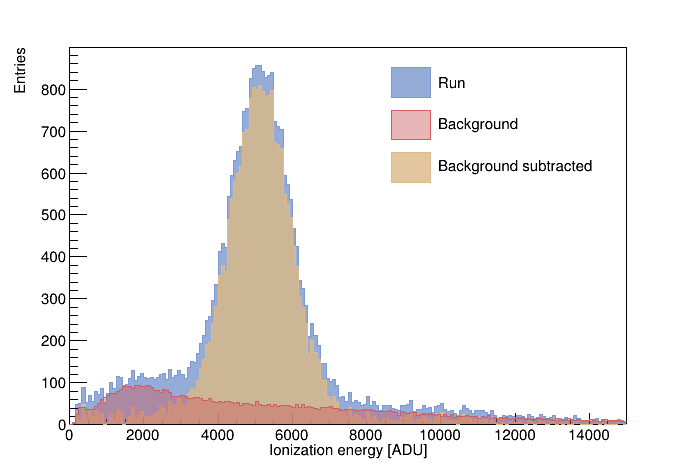}
    \caption{Background subtraction for $1.5~\mathrm{keV}$ electrons}
    \label{fig:bkgsub}
\end{figure}

The main background contribution comes from cosmic rays creating particles in the atmosphere, like muons, that can produce signals in the SPC in the keV-range \cite{cecchini}. The background remains constant throughout the experiment with a high signal-over-background ratio. In these conditions, the background energy spectrum, measured in the same experimental conditions as for Comimac measurements, can be subtracted from the other energy spectra after time normalization. The background rate has been measured to $23~\mathrm{Hz}$, a rate that remains low compared to the Comimac rate that varied between 50 Hz and 200 Hz depending on the particle species sent (electrons or ions, respectively). The proportion of Comimac events affected by random coincidences with muons is around $2\%$. Figure~\ref{fig:bkgsub} shows an example of the background subtraction for the lowest energy point exploited in the analysis ($1.5~\mathrm{keV}$ electrons).

Background subtraction and the correction of the ballistic deficit represent the only treatment of the data performed before the analysis. We do not apply any cut on our observables since well-resolved spectra were already obtained. 

\subsection{Energy lost in the interface between the SPC and Comimac}

The SPC is coupled with the Comimac beamline through a hole that has been drilled by a laser in a foil of stainless steel of thickness $13~\mathrm{\mu m}$. The ions and the electrons sent by Comimac pass through the hole and can lose part of their energy by scattering with the gas in the hole before reaching the instrumented part of the SPC. This energy loss must then be quantified in order to determine the kinetic energy of the particle when it enters the SPC volume. We characterized the quality of the hole with a Scanning Electron Microscope (SEM) at the NanoFab technological platform of the Néel institute in Grenoble. We measured a circular hole of diameter $1.45 \pm 0.06~\mathrm{\mu m}$.

\begin{figure}[H]
	\centering
	\includegraphics[width=\linewidth]{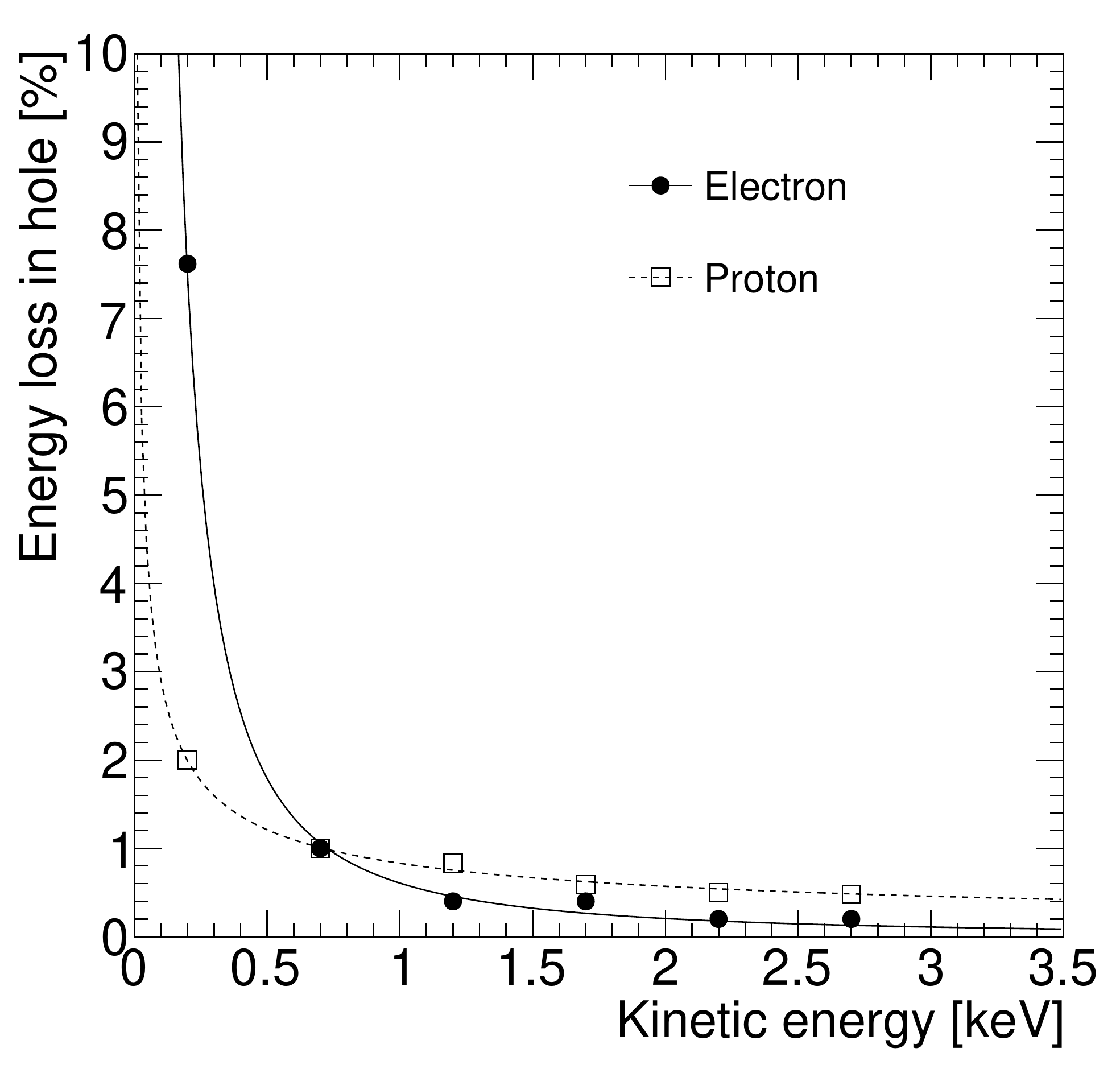}
	\caption{Simulation of the energy loss in the Comimac's hole of diameter $1.45~\mathrm{\mu m}$  for $100~\mathrm{mbar}$ of CH$_4$. The fit functions are displayed.}
	\label{fig:lossHole}
\end{figure}

With the hole properties in mind, we have performed simulations of the energy lost in the interface. The simulations are decomposed into two steps. Firstly, we perform a \texttt{Molflow+} \cite{Molflow+} simulation to determine the profile pressure inside the hole. \texttt{Molflow+} is a Monte Carlo simulation tool developed by CERN for high vacuum studies. Secondly, we use the obtained pressure profile as input for simulating the particle transport inside the hole. This second simulation is performed with \texttt{Casino} \cite{Casino} for the electrons and with \texttt{SRIM} \cite{SRIM} for the ions. \texttt{Casino} and \texttt{SRIM} are Monte Carlo simulation tools widely used for particle energies in the keV-range. In both cases, we set the initial kinetic energy of the particle and we compare it to the kinetic energy of the particle once it escapes the hole. This way we can approximate the amount of energy loss in Comimac's hole. The simulations also provide the energy distribution of the particle leaving the hole. When the energy loss is non-negligible, for instance when operating in some gas mixtures at atmospheric pressure, the energy spectra show a non-Gaussian tail similar to the ones in \cite{Cheng2016}. However, such tails are not observed in the data presented in this work, indicating that the energy loss in the interface is small.

The simulations show that the energy loss in the interface can be significant at large pressure ($\gtrsim 1~\mathrm{bar}$) for very low energies ($<0.5~\mathrm{keV}$). However, as presented in Figure~\ref{fig:lossHole}, for $100~\mathrm{mbar}$ of CH$_4$ the simulated energy loss in the interface hole remains below 1\% both for electrons and for protons in the considered energy range, \textit{i.e.} for $E_K > 1~\mathrm{keV}$. The percentage of energy loss obtained by simulations can be parametrized by a polynomial function $f(E_K) = \gamma~E_K^{-\delta}$ with $\gamma$ and $\delta$ determined by fitting the simulation results. In our analysis, we use this parametrization to correct the kinetic energy of the particles, \textit{i.e.} we subtract the amount of energy lost in the interface to the total kinetic energy given by Comimac. Such simulations can only give a first estimate of the energy loss, in particular since they rely on \texttt{SRIM} while we will later show some discrepancies between measurements and \texttt{SRIM} simulations. We thus decide to work with a conservative uncertainty of 100\% on the value of the energy loss. 

\subsection{Comparison between Comimac electrons and fluorescence photoelectrons}

We used the X-ray generator to produce photoelectrons of $1.49~\mathrm{keV}$ and $3.23~\mathrm{keV}$ \cite{ENSDF} by fluorescence as explained in the previous section. According to CXRO database \cite{CXRO}, 94\% of the fluorescence X-rays of the cadmium travel $15~\mathrm{cm}$ without interacting. For this reason, while this setup does not produce photoelectrons everywhere in the volume, it does produce them at all possible radii within the detector, and so we assume they will integrate any potential volume effects. On the contrary, the Comimac electrons are always sent at the same position: at the SPC surface through the hole. The range of the particles remains low: $7~\mathrm{mm}$ for $5~\mathrm{keV}$ electrons (according to \texttt{Casino}) and $1.4~\mathrm{mm}$ for $5~\mathrm{keV}$ protons (according to \texttt{SRIM}). Thus, comparing fluorescence photoelectrons and Comimac electrons is of particular interest for validating two assumptions: (1) that the electric field in the SPC is strong enough to collect all the charges, even from the surface; (2) that the kinetic energy of the electrons sent by Comimac is well determined. 

\begin{figure*}
	\centering
    \includegraphics[width=0.88\linewidth]{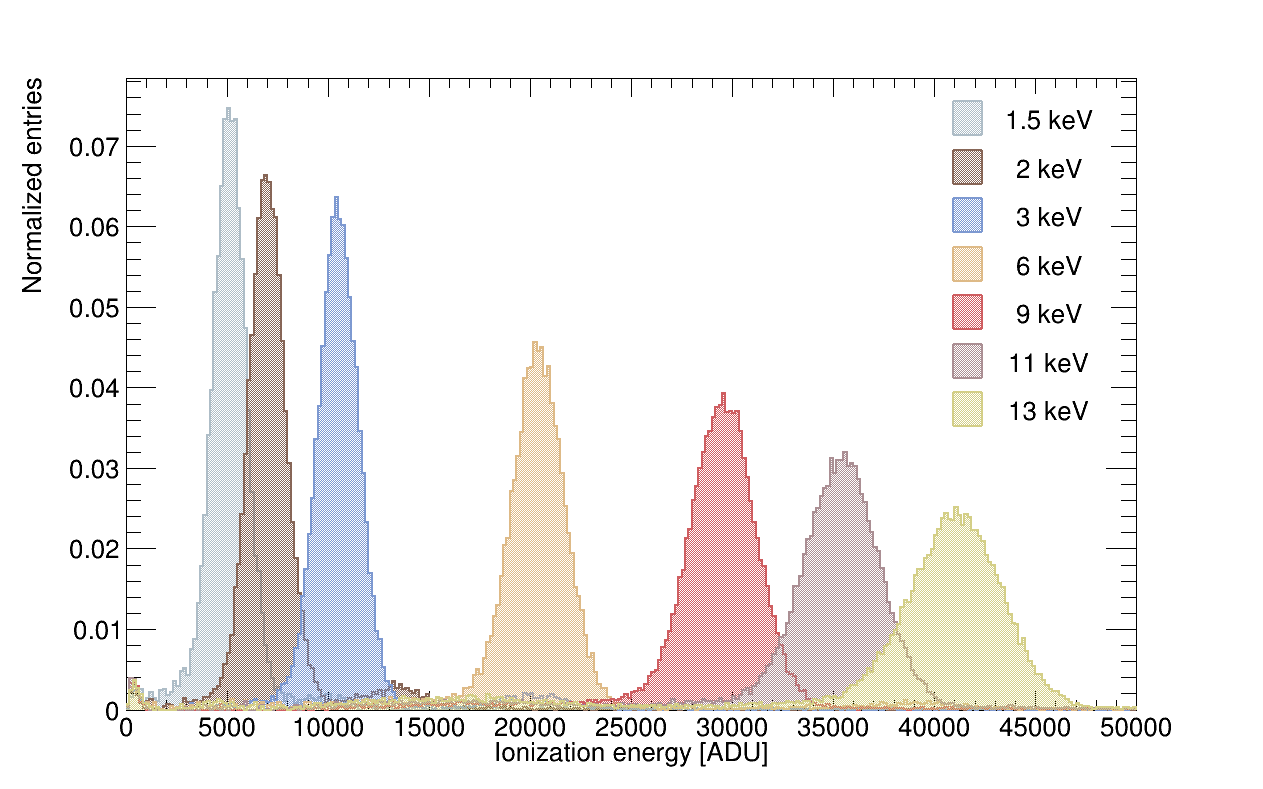}
    \caption{Complete set of energy spectra used for the calibration of the detector response. The kinetic energy is determined by the Comimac facility. The cosmic background has been subtracted but no cut is applied.}
    \label{fig:spectra}
\end{figure*}

We define $R \equiv (E_{\mathrm{Comimac}} - E_{\mathrm{X-rays}}) / E_{\mathrm{X-rays}}$, a ratio that quantifies the difference between the ionization energy measured for a fluorescence photoelectron and the one measured for a Comimac electron. We obtain $R(1.49~\mathrm{keV}) = 3.7\%~\pm~0.5\%$ and $R(3.23~\mathrm{keV}) = 3.9\%~\pm~0.5\%$, respectively for the aluminium line and for the cadmium line, indicating that we measure slightly more ionization energy for the Comimac electrons than for the photoelectrons, excluding energy losses in the interface between Comimac and the SPC as a possible explanation for the difference. The measurements with the X-ray generator have been performed with a fresh charge of gas compared to the other measurements but we applied the same working conditions. The different charge of gas has a non-negligible influence on the gain of the detector so we must scale the measured energies by comparing the position of the energy peaks in both datasets. This scaling brings an additional uncertainty that we propagate through our calculations.

From this analysis, we conclude that we efficiently collect the charges down to $1.49~\mathrm{keV}$, even the ones produced at the SPC surface and we validate the kinetic energy of Comimac. Due to the ACHINOS geometry, Comimac electrons mainly induce a signal on the anode ball aligned with the Comimac beam, while in the case of photoelectrons (having more heterogeneity in production position), a signal appears in several anode balls. Gain discrepancies between the anode balls \cite{Giomataris2020} could explain the measured differences, in particular since the energy resolution deteriorates by about $13\%$ for photoelectrons compared with Comimac electrons of the same energy. For the rest of the analysis, we will consider a conservative uncertainty of $4\%$ on the kinetic energy sent by Comimac. Note that this uncertainty also embeds the energy loss for electrons in the Comimac's hole described above.

The event rate with the X-ray generator is $\sim 400~\mathrm{Hz}$, which is about 8 times larger than for Comimac electrons. The measured energy could depend on the rate, for instance, if the ions drifting towards the cathode build a space charge \cite{Bohmer2012}. We have checked that in our conditions we are not sensitive to a rate effect. To do so, we have sent $15~\mathrm{keV}$ electrons with Comimac and we have varied the event rate up to $360~\mathrm{Hz}$ which is comparable to the rate obtained with the X-ray generator. We do not report any change in the measured energy up to this rate, from which we conclude that the X-ray measurements can be compared with the Comimac measurements. 

\section{Electron calibration}\label{sec:calib}

Electrons of known kinetic energies are sent in the detector by the Comimac facility. They are used to calibrate the detector by measuring the amplitude of the signal obtained for the Comimac electrons.

\begin{figure*}
	\centering
    \includegraphics[width=0.8\linewidth]{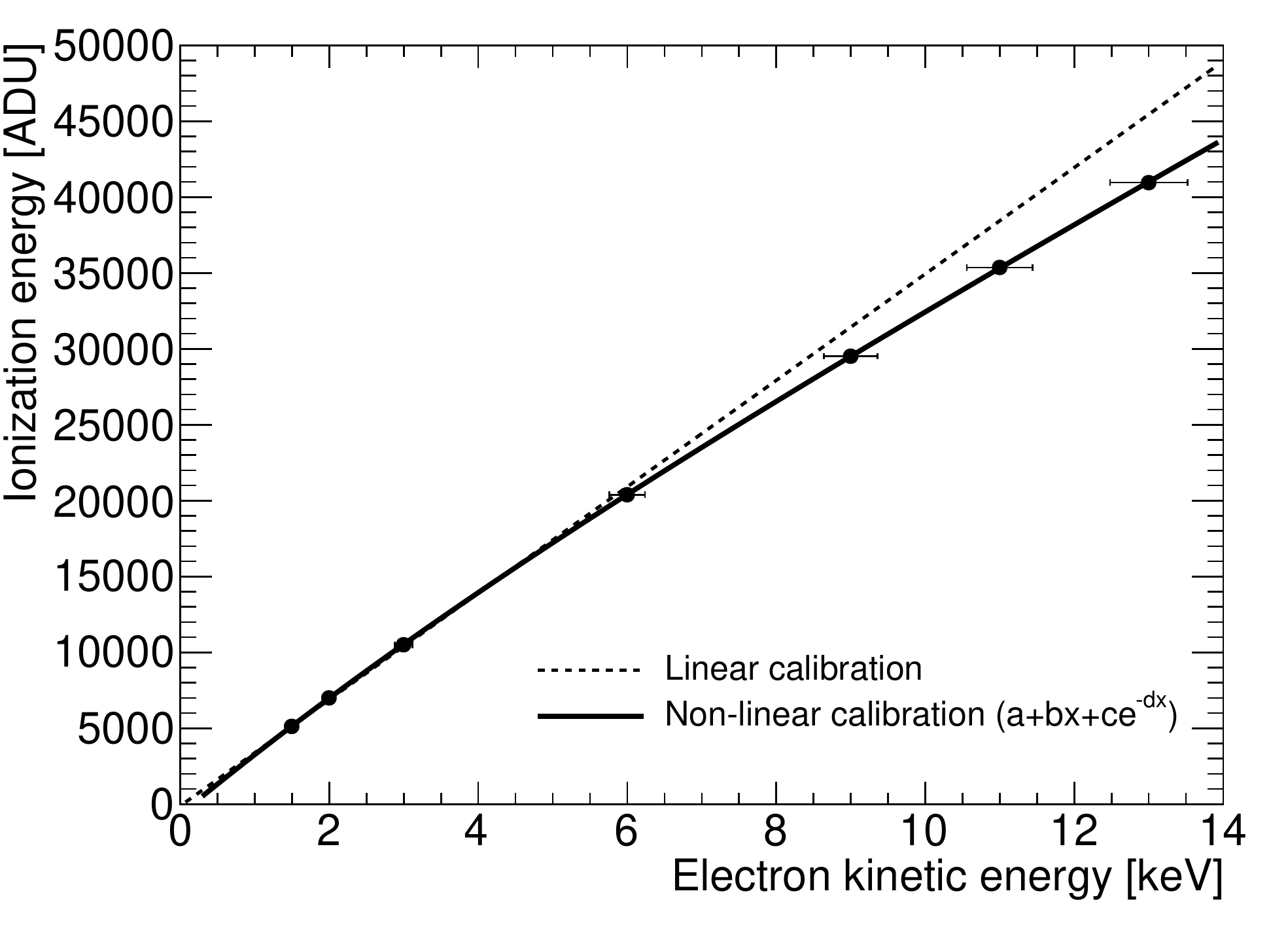}
    \caption{Electron calibration. The dashed line represents a linear calibration passing through the first data point and having an offset of $-117~\mathrm{ADU}$. The solid line is a fit with a first order polynomial function plus a decreasing exponential function. Error bars are drawn in X and Y but they are hardly visible in Y.}
    \label{fig:calibration}
\end{figure*}

Figure~\ref{fig:spectra} shows a superposition of the entire set of energy spectra obtained from Comimac electrons runs, after the background subtraction detailed in Section~\ref{sec:syst}. Each of the energy spectra presents the expected Gaussian shape and fits can be performed. The energy resolution, FWHM divided by the mean value, ranges between $[13.2~\%~,~ 32.7~\%]$. This good energy resolution, as well as the unambiguous determination of a Gaussian peak in each spectrum, leads to small statistical uncertainties compared to systematic uncertainties. The charge-to-voltage conversion of the CSP is determined by the use of a pulse generator combined with a capacitor to inject a charge into the CSP. Applying this conversion to the electron calibration, we measure an avalanche gain of $1.5\times10^4$ using a W-value of $27~\mathrm{eV}$ \cite{krajcar_bronic_mean_1988}. 

The electron calibration is shown in Figure~\ref{fig:calibration} taking into account the correction of the energy loss in the Comimac hole as detailed in Section~\ref{sec:syst}. The statistical uncertainties of the energy spectra are included in the error bars in Y. The error bars in X correspond to the uncertainty on the Comimac energy that embeds the $4\%$ difference observed with X-rays and the uncertainty in the energy loss in Comimac's hole.

The calibration presents a non-linear tendency that can be well-fitted by a first order polynomial function plus a decreasing exponential function. We have chosen to represent a linear function on the same plot, as a dashed line, this function passing through the first data point and having a small offset ($-117~\pm~15~\mathrm{ADU}$ which is equivalent to $-32~\mathrm{eV}$). This offset value corresponds to a linear extrapolation of the response of our acquisition chain. For this measurement, the SPC was coupled to a pulse generator and we varied the pulse amplitude to cover a detector response region from $3\times10^3~\mathrm{ADU}$ up to $26\times10^3~\mathrm{ADU}$. In other words, the dashed line in Figure~\ref{fig:calibration} represents the calibration expected from the response of the acquisition chain. This linear function highlights a departure from linearity above $4~\mathrm{keV}$ that could be due to the high gain of the sensor. While the exact mechanism that introduces this non-linearity is not fully understood, it has already been observed in Proportional Counters operating at high gain \cite{Srdoc,IonizationYield}. This phenomenon could possibly be explained by avalanche ions created by earlier primary electrons screening the avalanche field for later primary electrons, lowering the effective gain. The validation of this hypothesis requires further investigations out of the scope of the present work.

We have reproduced our measurements at a lower ACHINOS voltage, $1230~\mathrm{V}$ instead of $1270~\mathrm{V}$, in order to better understand the influence of the high gain on the departure from linearity. The ratio of the ionization energies at the two voltages remains constant in the entire tested region (between $3~\mathrm{keV}$ and $13~\mathrm{keV}$) with variations of less than 2\%. This analysis indicates that we are operating in a voltage region where the deviation from linearity seems not to depend on the avalanche gain.

\section{Ionization Quenching Factor}\label{sec:results}

\begin{figure*}
\centering
\includegraphics[width=\linewidth]{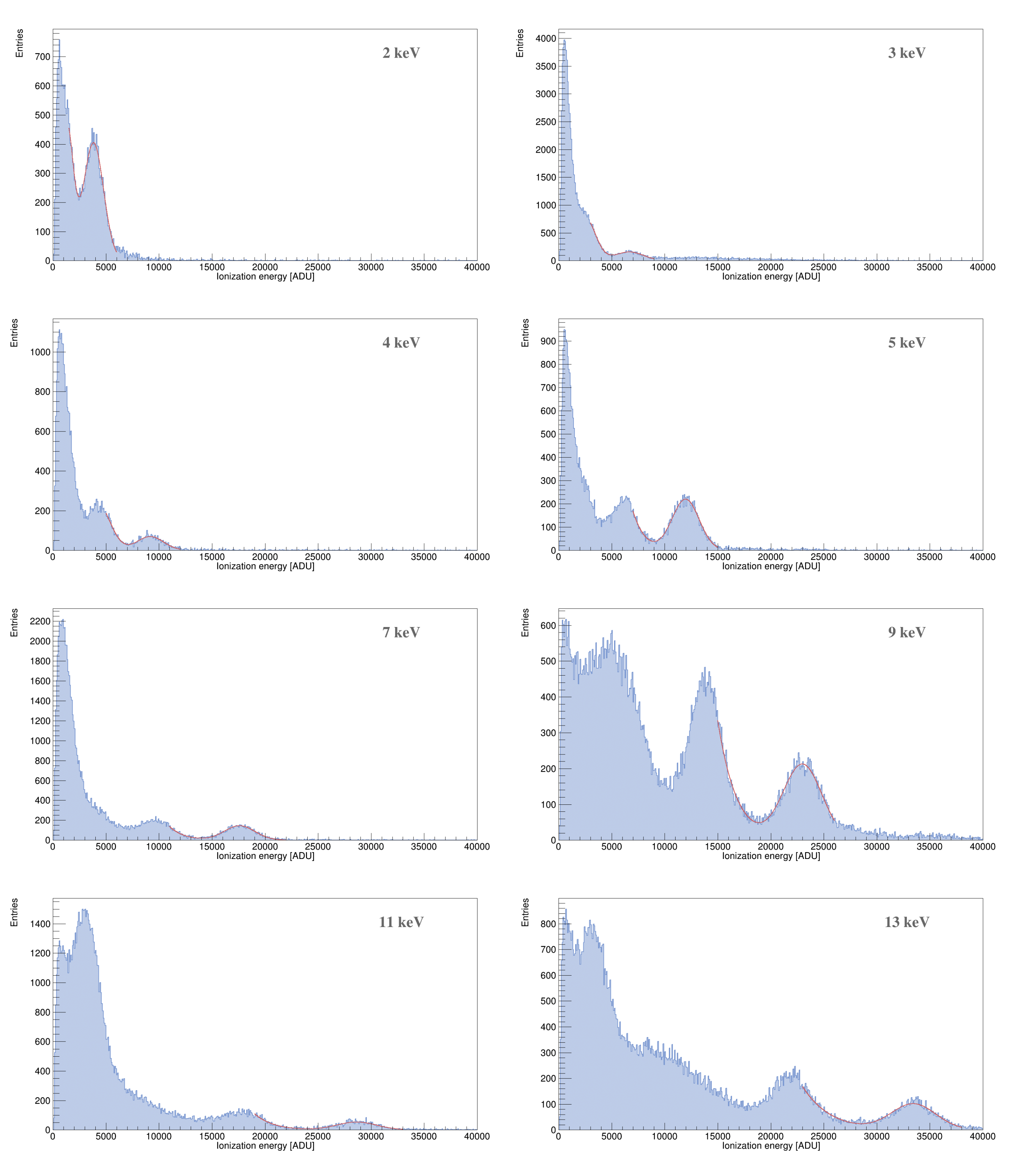}
	\caption{Complete set of proton spectra measured at $1270~\mathrm{V}$. The fit functions of two Gaussian, used for the analysis, are shown in red. In each spectrum, the proton peak can be identified as the one with the highest ionization energy. The second peak from the right corresponds to helium.}
	\label{fig:protons}
\end{figure*}

Protons of known kinetic energy are sent in the SPC by applying a positive extraction voltage on Comimac. The gas placed in the Comimac source, which turns into a plasma, contains a mixture of H$_2$ and He. For this reason, Comimac will send several types of ions: H$^+$, He$^+$, but also heavier species as N$^+$ or H$_2$O$^+$ due to water contamination and surface out-gassing in the source \cite{Comimac}. According to the Lindhard theory \cite{Lindhard}, the proton is the particle that releases the largest proportion of its kinetic energy through ionization. We thus identify the proton as the peak with highest ionization energy, the peak being Gaussian and well separated from the others. The complete set of measured proton spectra is presented in Figure~\ref{fig:protons}. We use a sum of two Gaussians as a fit function, modelling the proton and the helium peaks, with a fitting range starting in the tail of the helium peak in order to account for possible overlap between the peaks. In each spectrum, the relative amplitude of the peaks depends on the focusing parameters of the Comimac beam.

\begin{figure*}
\centering
\includegraphics[width=0.88\linewidth]{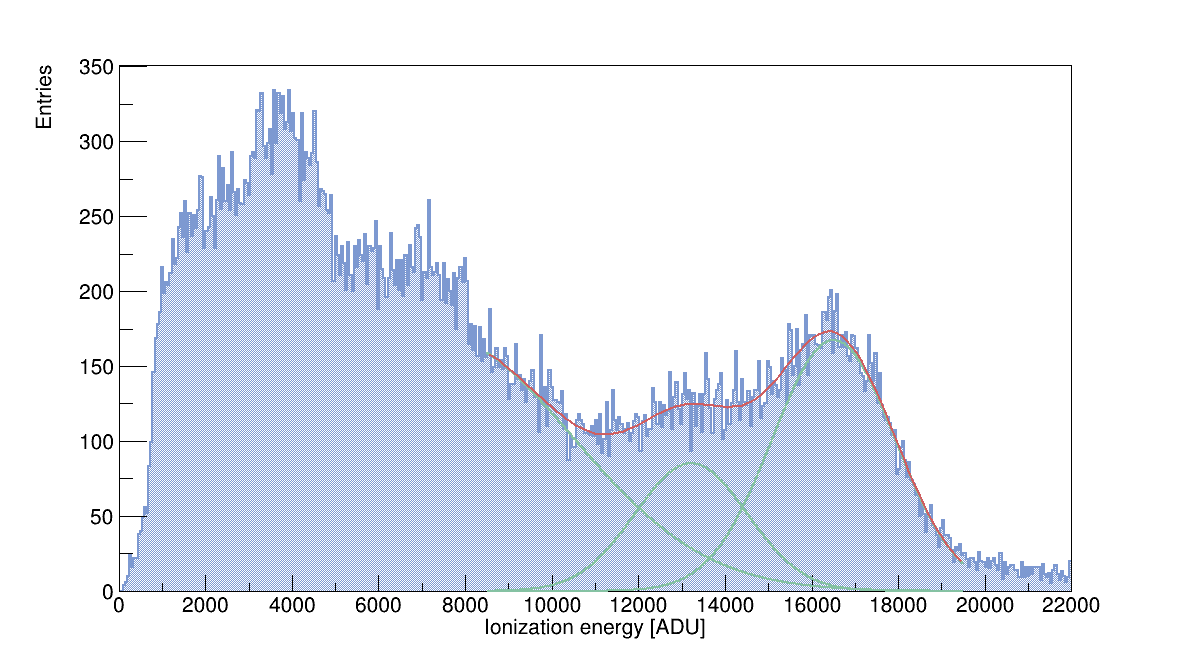}
	\caption{Example of a $9~\mathrm{keV}$ proton spectrum at $1230~\mathrm{V}$ showing the difficult separation between H$_2^+$ and H$^+$. The H$_2^+$ peak can be identified around $13000~\rm{ADU}$ whereas the H$^+$ peak is around $17000~\rm{ADU}$. The presence of H$_2^+$ is related to the low amount of He introduced in the Comimac source during this campaign.}
	\label{fig:H2separation}
\end{figure*}

The breaking of the H$_2$ molecule into protons in the Comimac source depends on the amount of He in the source: it must contain enough He so that the electron cyclotron resonance (ECR) sufficiently ionizes the plasma to break the molecular hydrogen. In the experiment performed at $1230~\mathrm{V}$, we observed the presence of an additional ion that we identified as H$_2^+$ as presented in Figure~\ref{fig:H2separation}. Its separation from the H$^+$ peak was only possible for kinetic energies above $9~\mathrm{keV}$. The presence of the H$_2^+$ contamination is not correlated with the gain of the SPC but only with the amount of He introduced in the Comimac source. While the contamination with H$_2^+$ in the source limited our IQF measurements at $1230~\mathrm{V}$, we do not observe this additional peak in the $1270~\mathrm{V}$ campaign. Note also that the IQFs for protons agree between the two campaigns, at least down to $9~\mathrm{keV}$. We thus conclude that we have efficiently broken the H$_2$ molecule into protons during the $1270~\mathrm{V}$ campaign, which is the one we use later on to parametrize the IQF.

We use the electron calibration as a reference for the kinetic energy, allowing us to rewrite Eq.~(\ref{eq:IQFdef}) as:
\begin{equation}
	\mathrm{IQF}~(E_K) ~ = ~  \frac{f_{\mathrm{calib}}(E_{\mathrm{ADU}})}{E_K} 
	\label{eq:IQFcalib}
\end{equation}
where $E_K$ is the ion kinetic energy and $f_{\mathrm{calib}}(E_{\mathrm{ADU}})$ is the calibration function that converts the measured ionization energy, $E_{\mathrm{ADU}}$, into the kinetic energy that an electron would require to produce such a signal. The main advantage of this approach is to compare directly the electrons and the ions produced in the same conditions. 

\begin{figure*}
	\centering
	\includegraphics[width=0.88\linewidth]{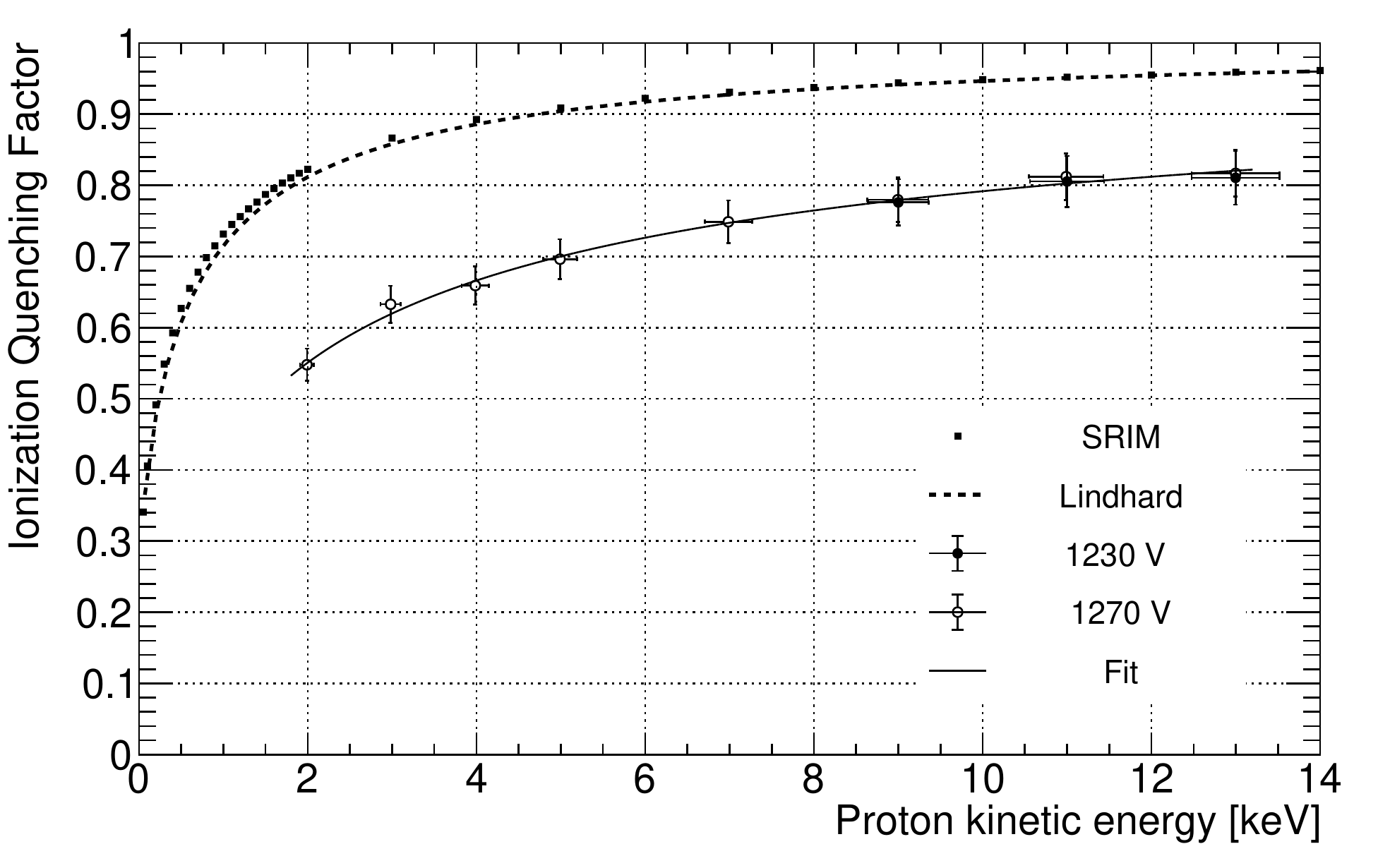}
	\caption{Ionization Quenching Factor for protons in $100~\mathrm{mbar}$ of methane. The measurements at $1230~\mathrm{V}$ and $1270~\mathrm{V}$ are respectively presented with black dots and white dots. Comparisons with \texttt{SRIM} and with the Lindhard theory are also shown.}
	\label{fig:IQF}
\end{figure*}

The measured IQF is presented in Figure~\ref{fig:IQF}. For comparisons, we show \texttt{SRIM} simulations and the prediction of the Lindhard theory with the parametrization described in \cite{Mei}. Measurements at $1230~\mathrm{V}$ and $1270~\mathrm{V}$ are presented and they agree, at least down to $9~\mathrm{keV}$. We consider this agreement as evidence that the reported IQF values do not depend on the observed high-gain effects, although definite proof would require data with larger spreads in gain. The uncertainties account for all the effects described in Section \ref{sec:syst} as well as statistical uncertainties. The main contribution to the error bar is due to the 4\% uncertainty on the Comimac kinetic energy. There was no calibration data for the ion recoils, so we assume that the uncertainty on the energy of the ions is the same as for the electrons. Following the approach of \cite{Mei} we propose a simplified parametrization of the IQF for the proton:
\begin{equation}
	\mathrm{IQF}~(E_K)~=~ \frac{E_K^\alpha}{\beta + E_K^\alpha}
\end{equation}
where we determine $\alpha = 0.70\pm0.08$ and $\beta = 1.32\pm0.17$ by fitting the experimental data. This fit function is also shown in Figure~\ref{fig:IQF}. To the authors' knowledge, this work presents the first measurement of the IQF of protons in methane.

We observe a deviation from \texttt{SRIM}, in particular at low energy, that reaches 33\% at $2~\mathrm{keV}$. Similar deviations with lower ionization energy than predicted by \texttt{SRIM} have already been observed in several gas mixtures \cite{Olivier,IQFSantos,Tampon,TheseQuentin,TheseDonovan}.  We recently measured a deviation from \texttt{SRIM} in the opposite direction (measuring more ionization energy) for neon nuclei in neon gas \cite{Marie} with a different setup: instead of Comimac, we used a neutron field to produce nuclear recoils whose energies are reconstructed from the measurement of their scattering angle. The \texttt{SRIM} code computes the ionization energy by following the approach of the Lindhard theory but using advanced calculations of the interatomic potentials based on a solid-state model in order to determine a \textit{universal} screening function that has been compared to hundreds of experimental data \cite{SRIM}. While this method remains the most precise on a general basis, it has been developed for higher energies $(\sim \mathrm{MeV / amu}$) than those treated in this work and some deviations and uncertainties have been pointed out for similar models in the keV-range. Also, it has to be noted that \texttt{SRIM} and the Lindhard theory do not take into account the gas density when computing the ionization energy, while we have previously observed that the IQF  in the keV-range depends also on the gas pressure \cite{IQFSantos}. 

\section{Conclusion}

The IQF represents a key parameter for low-mass WIMP searches through detection of ionization energy. The measurement of IQF is usually performed using neutron sources \cite{IQF_Ge,IQF_CDMS,IQF_CSI,Izraelevitch_2017,IQF_Ar,Marie} requiring coincidence detection under a high gamma ray background. The Comimac facility is an appropriate and convenient tool for IQF measurements in gases for its precision and its simplicity. It mimics the motion of a neutral atom with no need for neutron sources to induce nuclear recoils of known kinetic energies. Moreover, changing the polarity of Comimac enables to calibrate the detector with many electrons data points which, in turn, can be cross-referenced with known fluorescence peaks. The direct comparison between ions and electrons leads to the determination of the IQF. A potential disadvantage of the Comimac method, the ions being sent only at the surface instead of the whole volume, was accounted for by limiting the analysis above $1.5~\mathrm{keV}$, where full charge collection was demonstrated through volume calibrations with fluorescence photoelectrons. In the future, additional sources of photoelectrons will be used to access lower energies.

In this work, we dedicated special care to the study of systematic effects in order to validate the procedure and to quantify systematic uncertainties that propagate to the results. This work brought us to two main conclusions.

First, the calibration of the detector response presents a deviation from linearity above $4~\mathrm{keV}$ that we interpret as a high-gain effect as already observed in some Proportional Counters \cite{Srdoc,IonizationYield}. This non-linearity is taken into account in this work through the direct comparison of electronic and nuclear recoils in the same gas conditions. The phenomena responsible for the non-linearity could appear in proportional counters used in other experiments so we highlight the need for a calibration with multiple energy points. 

The second main conclusion concerns the significant discrepancy between the measured IQF and \texttt{SRIM} simulations since we measure less ionization energy than simulated, as also noted in previous works \cite{Olivier,IQFSantos,Tampon,TheseQuentin,TheseDonovan}. \texttt{SRIM} is commonly used to determine the IQF in gases for direct DM searches \cite{DMTPC_SRIM,Miuchi_SRIM}, however it has known limitations that impact the physics sensitivity of such experiments. This work present a method of directly measuring the IQF, removing this uncertainty, and providing an invaluable tool for DM searches. 

\newpage
\bibliographystyle{JHEP}
\bibliography{biblio}

\end{document}